\documentclass[12pt]{article}
\usepackage[export]{adjustbox}
\usepackage{graphicx}
\usepackage{graphics}
\usepackage{epsfig}
\usepackage{amssymb}
\usepackage{amsmath}
\usepackage{color}
\usepackage{textcomp}
\usepackage{latexsym}
\usepackage{physics}

\usepackage[margin=2.5cm,footskip=0.6cm]{geometry}
\begin{document}

\begin{center}
{\Large{\bf Velocity Distribution and Diffusion of an Athermal Inertial Run-and-Tumble Particle in a Shear-Thickening Medium}} \\
\ \\
by \\
Subhanker Howlader, Sayantan Mondal and Prasenjit Das\footnote{prasenjit.das@iisermohali.ac.in} \\
Department of Physical Sciences, Indian Institute of Science Education and Research -- Mohali, Knowledge City, Sector 81, SAS Nagar 140306, Punjab, India. \\
\end{center}

\begin{abstract}
\noindent We study the dynamics of an athermal inertial run-and-tumble particle moving in a shear-thickening medium in $d=1$. The viscosity of the medium is represented by a nonlinear function $f(v)\sim\tan(v)$, while a symmetric dichotomous noise of strength $\Sigma$ and flipping rate $\lambda$ models the activity of the particle. Starting from the Fokker-Planck~(FP) equation for the time-dependent probability distribution $W_{\pm\Sigma}(v, t)$ of the particle's velocity $v$  at time $t$ and the active force is $\pm\Sigma$, we analytically derive the steady-state velocity distribution function $W_s(v)$ and a quadrature expression for the effective diffusion coefficient $D_{\rm eff}$. For a fixed $\Sigma$, $W_s(v)$ undergoes multiple transitions with varying $\lambda$, and we have identified the corresponding transition points. We then numerically compute $W_s(v)$, the mean-squared velocity $\langle v^2\rangle(t)$, and the diffusion coefficient $D_{\rm eff}$, all of which show excellent agreement with the analytical results in the steady-state. Finally, we test the robustness of the transitions in $W_s(v)$ by considering an alternative $f(v)$ function that also capture the shear-thickening behavior of the medium.
\end{abstract}

\newpage
\section{Introduction}

\label{sec1}
Active matter comprises self-propelled particles that convert energy into directed motion, giving rise to intricate non-equilibrium phenomena at both the single-particle and collective scales\cite{S10,MJSTJR13,JGMJC22}. Examples of active matter are abundant in nature, ranging from microscopic entities such as bacterial colonies, motile cells, and synthetic microswimmers to macroscopic objects like bird flocks, fish schools, and human crowds~\cite{FTAS97,CRHCG16,AKTP21,DYGMYW19,J24,AF23}. In contrast to its passive counterpart, an active matter system breaks the fluctuation-dissipation relation and does not satisfy the detailed balance at the constituent level~\cite{EJ19}. Research in active matter has garnered significant interest in recent years for its ability to capture the complex dynamics of active systems, ranging from biological cells to engineered nanomotors~\cite{DZ17,I18,AAAG24}. Synthetic active systems, such as Janus particles, living crystals, and swarming robots, have opened new frontiers in materials science, smart drug delivery, autonomous robotics, and the design of self-organizing functional materials~\cite{TGA20,MYTJ20,DWU22}.

The two widely studied paradigms describing the self-propelled motion of active system constituents are Run-and-Tumble Particles (RTPs) and Active Brownian Particles (ABPs), each highlighting different aspects of active matter dynamics~\cite{PMWBL12,USAG18,IUS20}. The primary distinction between these models lies in their orientational dynamics: RTPs abruptly change their orientation through stochastic tumbling events between ballistic runs, whereas ABPs undergo smooth directional changes driven by rotational diffusion while maintaining a constant self-propulsion speed. Both models capture key features of active systems, including motility-induced phase separation (MIPS), persistent random walks, anomalous diffusion, and non-equilibrium pressure\cite{JM08,AYAMYMJ15,AMJ15,MKOMH16,SP24}. Recently, the Active Ornstein-Uhlenbeck Particle (AOUP) model has been proposed as a minimal description of active matter, incorporating noise characterized by an Ornstein-Uhlenbeck process to account for the persistence of self-propelled motion~\cite{GL30,YM12,CMNALD14,ECMJPF16}.

In most previous studies, the aforementioned models have been considered in the overdamped limit, where inertia is negligible and does not affect their motion. However, recent technological advancements have led to the development of self-propelled particles that are larger in size, higher in mass, or operate in low-viscosity environments, where inertia becomes more significant~\cite{H20}. Such inertia-dominated active particles are often referred to as ``microflyers", or ``runners" rather than microswimmers, as their underdamped dynamics resemble flying or gliding motion rather than swimming in highly viscous environments. Inertial active matter is widespread in nature, with flying insects, birds, and fish as natural examples, while vibrobots, microrobots, and self-propelled grains serve as experimental realizations~\cite{CH05,MG13,AALITASL0EM14,MFJG18,0V19}. These natural and experimental realizations have sparked a surge of theoretical studies on active particles with inertia~\cite{GRH22,LRH24,NM24,ALACH24}.

A cornerstone of non-equilibrium statistical mechanics is to understand the dynamics of an inertial passive particle moving through a fluid at a finite temperature~\cite{C85,H96,V08}. In the conventional framework, the fluid is typically treated as Newtonian, where the viscous force acting on the particle is linearly proportional to its velocity. However, many complex fluids found in nature—such as colloidal suspensions, polymer solutions, micellar fluids, paints, and dense granular materials—exhibit non-Newtonian behaviors like shear-thinning and shear-thickening, deviating from the linear response observed in simple Newtonian fluids~\cite{ANFGD08,J09,TA23,MD25}. In particular, for shear-thickening fluids, where the viscous force to motion increases with velocity, a linear viscous force becomes inadequate, necessitating nonlinear friction models that account for the velocity-dependent viscous force at high velocities. Recently, Menzel \textit{et al.} modeled shear-thickening using a nonlinear function, $f(v)\propto\tan(\gamma v)$, to study the dynamics of passive particles~\cite{A15}. At small velocities, $f(v)\sim \gamma v$, like that of a Newtonian fluid. However, at higher speeds, $f(v)$ increases nonlinearly and eventually diverges, imposing an upper limit on the particle's velocity. This behavior is a hallmark of shear-thickening environments, where the viscous resistance rises sharply with increasing particle velocity. An additional advantage of using $\tan(\gamma v)$ is that it remains smooth and differentiable throughout the entire range of allowed particle velocities.

The dynamics of an inertial run-and-tumble particle in a Newtonian fluid have been extensively studied in both the absence and presence of an external potential~\cite{DASU24,DAU24}. However, the motion of an inertial run-and-tumble particle in a shear-thickening medium remains poorly understood. In this paper, we explore the dynamics of an inertial active run-and-tumble particle moving through a shear-thickening medium in $d=1$. Many physical and biological systems exhibit effectively one-dimensional motion, such as particle transport through nanopores, optical traps, and along cytoskeletal filaments. Thus, our study will capture these scenarios, making it both realistic and physically well-motivated~\cite{NKA10,BLLRVV16}. We first adopt an approach similar to that of Menzel \textit{et al.}, where the viscous drag from the medium is modeled using a tangent-type function $f(v)\sim \tan(v)$. Understanding motion of active particles in shear-thickening media is crucial for biomedical applications like drug delivery and infection control, where microbes navigate complex bodily fluids. It also plays a key role in food safety and environmental microbiology, where bacteria move through viscoelastic materials~\cite{SD74,BT79,LWWH09,KGS23}. We represent the activity of the particle by a dichotomous Markov noise. To examine whether the results depend on the specific friction model, we also consider an alternative viscous drag force of the form $f(v)=v^n$, where $n>1$ is an odd integer. To the best of our knowledge, this is the first study of its kind in the literature.

Given this background, the structure of the paper is as follows: In Sec.~\ref{sec2}, we provide the details of the model and present the analytical calculations for the steady-state velocity distribution function, $W_s(v)$ and the effective diffusion coefficient, $D_{\rm eff}$. The numerical validation of our analytical results is discussed in Sec.~\ref{sec3}. In Sec.~\ref{sec4}, we discuss the robustness of the transitions observed in $W_s(v)$ by selecting an alternative function that also describes shear-thickening. Finally, we summarize our results in Sec.~\ref{sec5}.

\section{Model and Analytical Results}\label{sec2}
We consider the dynamics of an athermal inertial active particle moving through a shear-thickening medium. The viscous force experienced by the particle is modeled by a nonlinear function $f(v)\sim \tan(v)$. The Langevin equation for the particle's motion is given by
\begin{eqnarray}
\label{apeqn1}
m \frac{d\vec{v}}{dt} = -A\tan\left(\frac{|\vec{v}|}{a}\right) \hat{v} + \vec{\xi}(t).
\end{eqnarray}
In Eq.~(\ref{apeqn1}),  $m$ and $\vec{v}$ denote the mass and velocity of the particle, respectively, while $\hat{v}$ represents the unit vector along $\vec{v}$. The first term on the right-hand side, $-A \tan\left(|\vec{v}|/a\right) \hat{v}$, represents the nonlinear viscous force, where $A$ sets the strength of the viscous force and $a$ defines a characteristic velocity scale. The term $\vec{\xi}(t)$ represents the active force, modeled as a symmetric dichotomous noise~\cite{WR83,J84,I06}.

For simplicity, we consider Eq.~(\ref{apeqn1}) in $d=1$. In this case, the equation of motion reduces to
\begin{eqnarray}  
\label{apeqn1a}  
m \frac{dv}{dt} = -A \tan\left(\frac{v}{a}\right) + \xi(t).  
\end{eqnarray}
Here, the symmetric dichotomous noise $\xi(t)$ takes only two discrete values, $\pm\nu$, switching between them with a probability $\Gamma dt$ in time $dt$, where $\Gamma$ represents the uniform transition rate. It has a zero mean and satisfies the autocorrelation function
\begin{eqnarray}  
\label{apeqn2}  
\langle \xi(t) \xi(t^{\prime}) \rangle = \nu^2 \exp\left(-2 \Gamma |t - t^{\prime}|\right).  
\end{eqnarray}
This noise effectively models the run-and-tumble dynamics of living organisms in $d=1$.

Next, we nondimensionalize Eq.~(\ref{apeqn1a}) by introducing the following rescaled variables for time and velocity:
\begin{eqnarray}  
\label{apeqn2a}  
t = \frac{m a}{A} t^\prime \quad \text{and} \quad v = a v^\prime,  
\end{eqnarray}
where the primed variables represent the corresponding dimensionless quantities. Substituting these into Eq.~(\ref{apeqn1a}) and omitting the primes, we obtain its nondimensional form:  
\begin{eqnarray}  
\label{apeqn1b}  
\frac{dv}{dt} = -\tan(v) + \eta(t).  
\end{eqnarray}
The scaled active force $\eta(t)$ satisfies the following properties:  
\begin{eqnarray}  
\label{apeqn2b}  
\langle \eta(t) \rangle = 0  \quad \text{and} \quad \langle \eta(t) \eta(t^{\prime}) \rangle = \Sigma^2 \exp\left(-2\lambda|t - t^{\prime}|\right),  
\end{eqnarray}  
where $\Sigma = \nu / A$ and $\lambda = m a \Gamma / A$, respectively, represent the scaled strength and flipping rate of the active force.

We identify the stable fixed points of the deterministic velocity dynamics, which determine the maximum attainable velocity $v_m$ of the particle for a given state of $\eta(t)$. At these fixed points, we find:
\begin{eqnarray}  
\label{apeqn3}  
\tan\left(v_m\right) = \Sigma \quad \text{and} \quad \tan\left(v_m\right) = -\Sigma,  
\end{eqnarray}  
which gives  
\begin{eqnarray}  
\label{apeqn4}  
v_m = \tan^{-1}\Sigma \quad \text{and} \quad v_m = -\tan^{-1}\Sigma.  
\end{eqnarray}  
This implies that the particle’s velocity will always be confined within the range $v_m = \pm \tan^{-1} \Sigma$.

\subsection{Derivation of Steady State Velocity Distribution Function $W_s(v)$}\label{2a}
We consider $W_{+\Sigma}(v, t)$ as the probability density function describing the velocity $v$ of the particle at time $t$ and the active force is $+\Sigma$. Likewise, $W_{-\Sigma}(v, t)$ defines the probability density function for the velocity $v$ at time $t$, along with the active force is $-\Sigma$. The corresponding Fokker-Planck (FP) equations for these probability densities, derived from the dynamical equation (\ref{apeqn1b}), are given by
\begin{eqnarray}  
\label{apeqn5}  
&&\frac{\partial}{\partial t} W_{+\Sigma}(v, t) = -\frac{\partial}{\partial v} \left[ -\tan{\left(v\right)}+\Sigma  \right]W_{+\Sigma}(v, t) - \lambda\left[W_{+\Sigma}(v, t) - W_{-\Sigma}(v, t)\right], \\  
\label{apeqn6}  
&&\frac{\partial}{\partial t} W_{-\Sigma}(v, t) = -\frac{\partial}{\partial v} \left[ -\tan{\left(v\right)}-\Sigma  \right]W_{-\Sigma}(v, t) + \lambda\left[W_{+\Sigma}(v, t) - W_{-\Sigma}(v, t)\right].  
\end{eqnarray}\\
In Eqs.~(\ref{apeqn5}) and (\ref{apeqn6}), the first term on the right-hand side captures the deterministic evolution of the probability densities, driven by the forces $-\tan(v)$ and $\pm\Sigma$, which determine their redistribution across velocity space. The second term corresponds the stochastic transitions of the active force $\eta(t)$, facilitating the exchange of probability between the two force states at a given velocity $v$. Our focus is on the probability density $W(v,t)$, which represents the probability of the particle having velocity $v$ at time $t$, regardless of the active force. It is defined as $W(v,t) = W_{+\Sigma}(v, t) + W_{-\Sigma}(v, t)$. Additionally, we introduce a new function $U(v,t)$, given by $U(v,t) = \lambda [W_{+\Sigma}(v, t) - W_{-\Sigma}(v, t)]$. 

We recast Eqs.~(\ref{apeqn5}) and (\ref{apeqn6}) in terms of $W(v,t)$ and $U(v,t)$ as
\begin{eqnarray}  
\label{apeqn7}  
&&\frac{\partial}{\partial t} W(v,t) = -\frac{\partial}{\partial v} \left[ -\tan{\left(v\right)} \right]W(v,t) - \frac{\Sigma}{\lambda} \frac{\partial}{\partial v} U(v,t), \\  
\label{apeqn8}  
&&\frac{\partial}{\partial t} U(v,t) = -\frac{\partial}{\partial v} \left[ -\tan{\left(v\right)} \right] U(v,t) - 2\lambda U(v,t) - \Sigma \lambda \frac{\partial}{\partial v} W(v,t).  
\end{eqnarray}
To solve Eq.~(\ref{apeqn8}), we assume an initial condition based on a natural statistical consideration: in the distant past, the velocity \( v \) and the active force \( \eta \) were uncorrelated. This implies that \( U(v, t) \) satisfies \( U(v, -\infty) = 0 \). With this assumption, the solution to Eq.~(\ref{apeqn8}) takes the form
\begin{eqnarray}  
\label{apeqn9}  
U(v,t)=-\int^{t}_{-\infty}dt^{\prime} \exp\left[-\left\{2\lambda+\frac{\partial}{\partial v} \left(- \tan{\left(v\right)}\right)  \right\}(t - t^{\prime})\right] \Sigma\lambda  
\frac{\partial}{\partial v} W(v, t^{\prime}).  
\end{eqnarray}
Substituting this expression for \( U(v,t) \) into Eq.~(\ref{apeqn7}), we obtain a closed-form equation for \( W(v,t) \):
\begin{eqnarray}  
\label{apeqn10}  
\frac{\partial}{\partial t} W(v, t) &=& -\frac{\partial}{\partial v} \left[ -\tan{\left( v\right)}  \right]W(v,t) \nonumber \\
&& + \Sigma^2\frac{\partial}{\partial v} \int^{t}_{-\infty}dt^{\prime} \exp\left[-\left\{2\lambda+\frac{\partial}{\partial v} \left(- \tan{\left(v\right)}\right)  \right\}(t - t^{\prime})\right]  
\frac{\partial}{\partial v} W(v, t^{\prime}).  
\end{eqnarray}
Since Eq.~(\ref{apeqn10}) involves derivatives of \( W(v,t^{\prime}) \) of arbitrarily high order, solving for its complete time evolution is not straightforward. However, by imposing natural boundary conditions, we can obtain the stationary probability distribution \( W_s(v) \).

To compute \( W_s(v) \), we start with the steady-state versions of Eqs.~(\ref{apeqn7}) and (\ref{apeqn8}):\begin{eqnarray}  
\label{apeqn11}  
&& 0 = -\frac{\partial}{\partial v} \left[ -\tan{\left(v\right)} \right]W_s(v) - \frac{\Sigma}{\lambda}\frac{\partial}{\partial v} U_s(v), \\  
\label{apeqn12}  
&& 0 = -\frac{\partial}{\partial v} \left[ -\tan{\left( v\right)}\right]U_s(v) - 2\lambda U_s(v) - \Sigma\lambda\frac{\partial}{\partial v}W_s(v).  
\end{eqnarray}
From Eq.~(\ref{apeqn11}), we obtain
\begin{eqnarray}  
\label{apeqn13}  
\left[ -\tan{\left(v\right)} \right]W_s(v) + \frac{\Sigma}{\lambda} U_s(v)=C.  
\end{eqnarray}
Since the velocity dynamics in Eq.~(\ref{apeqn1b}) is deterministically stable, both $W_s(v)$ and $U_s(v)$ vanish as $|v|>v_m$. This implies that $C=0$. So the Eq.~(\ref{apeqn13}) gives
\begin{eqnarray}  
\label{apeqn14}  
U_s(v)=\frac{\lambda}{\Sigma}\left[\tan{\left(v\right)} \right]W_s(v).  
\end{eqnarray}
Substituting this expression for \( U_s(v) \) and its derivative into Eq.~(\ref{apeqn12}), we obtain
\begin{eqnarray}  
\label{apeqn16}  
W^{\prime}_s(v)&=&\frac{2\tan(v)}{\Sigma^2-\tan^2(v)}\left\{\sec^2(v)-\lambda\right\}W_s(v).  
\end{eqnarray}
Integrating Eq.~(\ref{apeqn16}) yields the stationary probability density function:
\begin{eqnarray}  
\label{apeqn17}  
W_s(v) = \frac{N}{\Sigma^2  - \tan^2(v)}  
\exp \left\{-2\lambda\int_{}^{v} dv'\frac{\tan(v')}{\Sigma^2 - \tan^2(v')} \right\}.  
\end{eqnarray}
We evaluate the integral in Eq.~(\ref{apeqn17}), which leads to the following stationary probability distribution for $v\in (-v_m,+v_m)$: 
\begin{eqnarray}  
\label{apeqn18}  
W_s(v) = \frac{N}{\Sigma^2 -\tan^2(v)}  
\left[ \frac{\Sigma^2- \tan^2(v)}{1 + \tan^2(v)} \right]^{\frac{\lambda}{(1+\Sigma^2)}}.  
\end{eqnarray}
The constant $N$ remains undetermined analytically, as the integral of $W_s(v)$ does not admit a closed-form solution. However, $N$ can be calculated numerically for all allowed values of $\lambda$ and $\Sigma$. Further, the expression for $W_s(v)$ explicitly exhibits symmetry under the transformation $v \rightarrow -v$. 
\begin{figure}
\centering
\includegraphics*[width=0.55\textwidth]{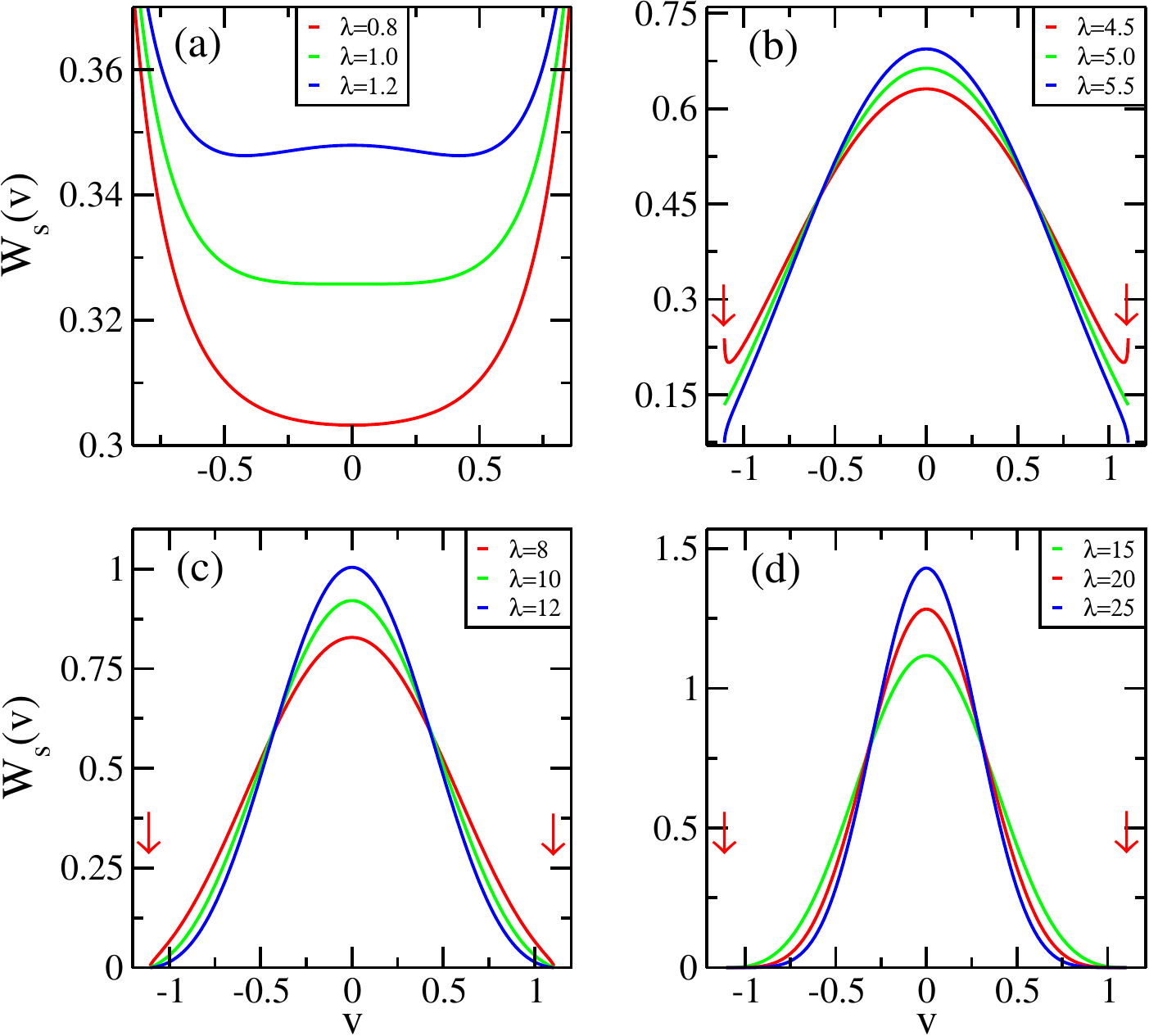}
\caption{\label{fig1} Plot of the normalized steady-state velocity distribution function $W_s(v)$ for $\Sigma=2.0$ with the following ranges of $\lambda$: (a) $\lambda\in(0.8,1.2)$, (b) $\lambda\in(4.5,5.5)$, (c) $\lambda\in(8,12)$, and (d) $\lambda\in(15,25)$, as indicated. The red arrows in panels (b)–(d) indicate the boundaries of the velocity domain.}
\end{figure}

In Fig.~\ref{fig1}, we plot the normalized $W_s(v)$ for $\Sigma = 2$ at various values of $\lambda$, as indicated. $W_s(v)$ exhibits multiple transitions as $\lambda$ varies. Specifically, $W_s(v)$ diverges at $v=\pm v_m$ and shows a minimum at $v=0$ when $\lambda<1$. However, when $\lambda>1$, a local maximum emerges at $v=0$ while the divergence at $v=\pm v_m$ remains unchanged, as shown in Fig.~\ref{fig1}(a). Here, an increase in $\lambda$ leads to more frequent flipping of active force $\eta(t)$, resulting in the occurrence of the peak at $v=0$. When $\lambda$ increases further, $W_s(v)$ undergoes a transition at $v = \pm v_m$, diverging to infinity for $\lambda < 5$ and approaching zero vertically for $\lambda > 5$, as shown in Fig.~\ref{fig1}(b). However, for $\lambda \gtrsim 5$, the particle velocity primarily fluctuates around $v=0$ and can still reach $v = \pm v_m$, leading to a rapid suppression of $W_s(v)$ at $v = \pm v_m$. For $8<\lambda<12$, the tails of $W_s(v)$ approaches zero vertically for $\lambda<10$ and horizontally for $\lambda>10$ at $v=\pm v_m$, as shown in Fig.~\ref{fig1}(c). Essentially, for $\lambda>10$, the particle is unable to reach the $v = \pm v_m$, causing $W_s(v)$ to drop to zero horizontally. As $\lambda$ increases further, $W_s(v)$ becomes a bell-shaped curve centered at $v=0$, with the peak height increasing with $\lambda$, as shown in Fig.~\ref{fig1}(d). These observations call for a deeper analysis of $W_s(v)$ in relation to the active force parameters $\Sigma$ and $\lambda$.

\subsection{Activity-Induced Transitions in $W_s(v)$}\label{2b}
We begin by analyzing the extrema of $W_s(v)$. Differentiating Eq.~(\ref{apeqn18}) with respect to $v$ gives the locations of the extrema at $v=u$, leading to the following relation:
\begin{eqnarray}
\label{apeqn23}
\tan(u) - \frac{1}{\lambda}\tan(u)\sec^2(u)=0.
\end{eqnarray}
Roots of the Eq.~(\ref{apeqn23}) are at
\begin{eqnarray}
\label{apeqn24}
u_0 = 0 \quad \text{and}\quad u_{1,2} = \pm  \tan^{-1}\left(\sqrt{\lambda-1}\right).
\end{eqnarray}
For $\lambda<1$, $v=u_0=0$ is the only physical solution of Eq.~(\ref{apeqn23}) and corresponds to a minimum of $W_s(v)$. However, for $\lambda>1$, two new minima of $W_s(v)$ emerge at $v=u_{1,2}$, while $v=u_0$ becomes a maximum. From Eq.~(\ref{apeqn24}), it is clear that the positions of the minima depend only on $\lambda$. These results are consistent with our observation shown in Fig.~\ref{fig1}(a).

Next, we examine the behavior of $W_s(v)$ at $v=\pm v_m$, the fixed points or attractors of velocity dynamics of the particle. Since $W_s(v)$ is symmetric about $v=0$, it is sufficient to analyze its behavior at a single fixed point, either $v=+v_m$ or $v=-v_m$. Here, we will focus primarily on the behavior at $v=+v_m$. The Taylor expansion of $W_s(v)$ around $v=+ v_m$ yields
\begin{eqnarray}
\label{apeqn19}
W_s(v) \sim \left(v_m -v\right)^{\frac{\lambda}{(1+\Sigma^2)}-1}.
\end{eqnarray}
$W_s(v)$ diverges to infinity at $v=+v_m$ in an integrable manner when
\begin{equation}
\label{apeqn20}
\frac{\lambda}{1 + \Sigma^2} < 1.
\end{equation}
This condition holds when $\lambda<1+\Sigma^2$. Since the amplitude of active force $\Sigma$ is always positive, $\lambda\leq 1$ will always cause $W_s(v)$ to diverge at $v=+v_m$. Thus, any non-zero $\Sigma$ extends the range of $\lambda$ for which $W_s(v)\rightarrow \infty$ at $v=+v_m$. When $\lambda$ satisfies $\lambda>1+\Sigma^2$, $W_s(v)$ approaches zero at $v=\pm v_m$ with either a divergent slope or a zero slope. In this regime the flipping dynamics of the active force will effectively suppress the influence of the attractors at $v=+v_m$ causing $W_s(v)$ to approach zero. The pair of minima that emerge for $\lambda>1$ persist up to $\lambda = 1+\Sigma^2$.

To examine the precise manner in which $W_s(v)\rightarrow 0$ at $v=+v_m$, we differentiate Eq.~(\ref{apeqn19}), which yields:
\begin{eqnarray}
\label{apeqn22}
W_s^{\prime}(v) \sim \left(v_m -v\right)^{\frac{\lambda}{(1+\Sigma^2)}-2}
\end{eqnarray}
near $v=+v_m$. Clearly, $W_s^{\prime}(v)\rightarrow\infty$ at $v=+v_m$ if $\lambda<2(1+\Sigma^2)$. Therefore, we can conclude that for the range $(1+\Sigma^2)<\lambda<2(1+\Sigma^2)$, $W_s(v)\rightarrow 0$ while $W_s^{\prime}(v)\rightarrow\infty$  at $v=+v_m$. This indicates a cusp behavior at $v=+ v_m$, where the probability density $W_s(v)$ decreases to zero with an infinite slope. Again, $W_s^{\prime}(v)\rightarrow 0$ at $v=+v_m$ if $\lambda>2(1+\Sigma^2)$. Therefore, for $\lambda>2(1+\Sigma^2)$, we observe that both $W_s(v)$ and $W_s^{\prime}(v)$ approaches to zero  at $v=+v_m$, i.e., $W_s(v)$ decreases to zero with zero slope.

Thus, the analysis of $W_s(v)$ under varying conditions reveals the following scenarios:\ \\
(a) $\lambda \leq1$: In this regime, we find that $W_s(v)$ has a minimum at $v=0$ while diverging to infinity at $v=\pm v_m$. \ \\
(b) $1<\lambda < 1 + \Sigma^2$: Here, $W_s(v)$ has a maximum at $v = 0$ is the maximum and a pair of minima at $v=\pm\tan^{-1}\left(\sqrt{\lambda-1}\right)$, while still diverging at $v=\pm v_m$. \ \\
(c) $(1 + \Sigma^2)<\lambda<2(1+\Sigma^2)$: In this case, $v = 0$ remains the maximum of $W_s(v)$, while $W_s(v)\rightarrow 0$ at $v = \pm v_m$ with a cusp behavior. \ \\
(d) $\lambda>2(1+\Sigma^2)$: In this range, both $W_s(v)$ and $W_s^\prime(v)$ tends to zero, while $v = 0$ remains the maximum of $W_s(v)$.\ \\
These findings align with the results shown in Fig.~\ref{fig1} for $\Sigma=2$, where the transition points are calculated as $\lambda_1=1$ in Fig.~\ref{fig1}(a), $\lambda_2=1+\Sigma^2=5$ in Fig.~\ref{fig1}(b), and $\lambda_3=2(1+\Sigma^2)=10$ in Fig.~\ref{fig1}(c). Since the transitions depend on the flipping rate $\lambda$ and the active force strength $\Sigma$, we refer to them as \textit{Activity-Induced Transitions}.

\subsection{Effective Diffusion Coefficient $D_{\rm eff}$}\label{sec2c}
To gain deeper insight into the transport behavior of an active particle in a shear-thickening medium, we evaluate the effective diffusion coefficient $D_{\rm eff}$. Using the Kubo relation,  $D_{\rm eff}$ is given by the integral of the velocity autocorrelation function (VACF) as follows:
\begin{eqnarray}
\label{apeqn25}
D_{\rm eff} = \int_0^\infty dt \left[\langle v(0)v(t) \rangle - \langle v(0) \rangle^2\right] 
\end{eqnarray}
in $d=1$. As the velocity distribution function is an even function of $v$, it directly implies that $\langle v(0) \rangle = 0$. As a result, Eq.~(\ref{apeqn25}) simplifies to  
\begin{eqnarray}  
\label{apeqn26}  
D_{\rm eff} = \int_0^\infty dt \, \langle v(0)v(t) \rangle.  
\end{eqnarray}
A quadrature expression for the diffusion coefficient of a particle driven by dichotomous noise and subject to a general odd friction function was previously derived by Lindner~\cite{B10}. Using a similar approach, we obtain the following expression for $D_{\rm eff}$:
\begin{eqnarray}
\label{apeqn27}
D_{\rm eff} = 2\lambda \Sigma^2 \frac{\int_0^{v_m} dx \, e^{V(x)} \left[\Sigma^2 - \tan^2(x) \right]^{-1}       \left\{\int_x^{v_m} dy \,y e^{-V(y)} \left[\Sigma^2 -  \tan^2(y) \right]^{-1}\right\}^2}{\int_0^{v_m} dz \, e^{-V(z)} \left[\Sigma^2 - \tan^2(z) \right]^{-1}},
\end{eqnarray}
where the function $V(x)$ has the following form:
\begin{eqnarray}
\label{apeqn28}
V(x) = 2\lambda \int_0^x dx^\prime \, \frac{\tan(x^\prime)}{\Sigma^2 - \tan^2(x^\prime)}.
\end{eqnarray}
Clearly, $D_{\rm eff}$ depends on the active force parameters $\Sigma$ and $\lambda$. Since the exact nature of this dependence cannot be determined analytically due to the absence of closed-form solutions for the integrals, we evaluate it computationally in Sec.~\ref{sec3}.

\section{Numerical Results}\label{sec3}
We numerically solve Eq.~(\ref{apeqn1b}) using the Euler-discretization method to update the particle's velocity and compute the mean-squared velocity $\langle v^2\rangle(t)$, steady-state velocity distribution function $W_s(v)$, and velocity auto-correlation function $\langle v(0)v(t)\rangle$ in $d=1$. The discretized version of Eq.~(\ref{apeqn1b}) is given by:
\begin{eqnarray}
\label{apeqn31}
v(t+dt)=v(t) - [\tan v(t) - \eta(t)] dt.
\end{eqnarray}
We denote the position of the particle at time $t$ as $r(t)$, which is updated using the following discretized equation:
\begin{eqnarray}
\label{apeqn32}
r(t+dt)=r(t) + v(t) dt.
\end{eqnarray}
Here, $dt$ is the discretized time step. We generate the dichotomous noise $\eta(t)$ following the method outlined in Ref.~\cite{CE06}. The probability $p$ that $\eta(t)$, starting from state $+\Sigma$(or $-\Sigma$), transitions to state $-\Sigma$(or $+\Sigma$) in time interval $dt$ is given by  
\begin{eqnarray}
\label{apeqn33}
p=\frac{1}{2}\{1-\exp(-2\lambda dt)\}.
\end{eqnarray}
Next, we generate a random number $w$ in the interval [0,1] using a uniform random number generator. When $p>w$, $\eta(t)$ transitions from one state to another. In the simulation, We set the numerical value of $dt=0.001$ and all the statistical results are obtained by averaging over $10^5$ trajectories. At $t=0$, a particle is initialized with zero velocity and positioned at the origin. We update the velocity and position of the particle iteratively using Eqs.~(\ref{apeqn31}) and (\ref{apeqn32}), respectively.

\begin{figure}
\centering
\includegraphics*[width=0.5\textwidth]{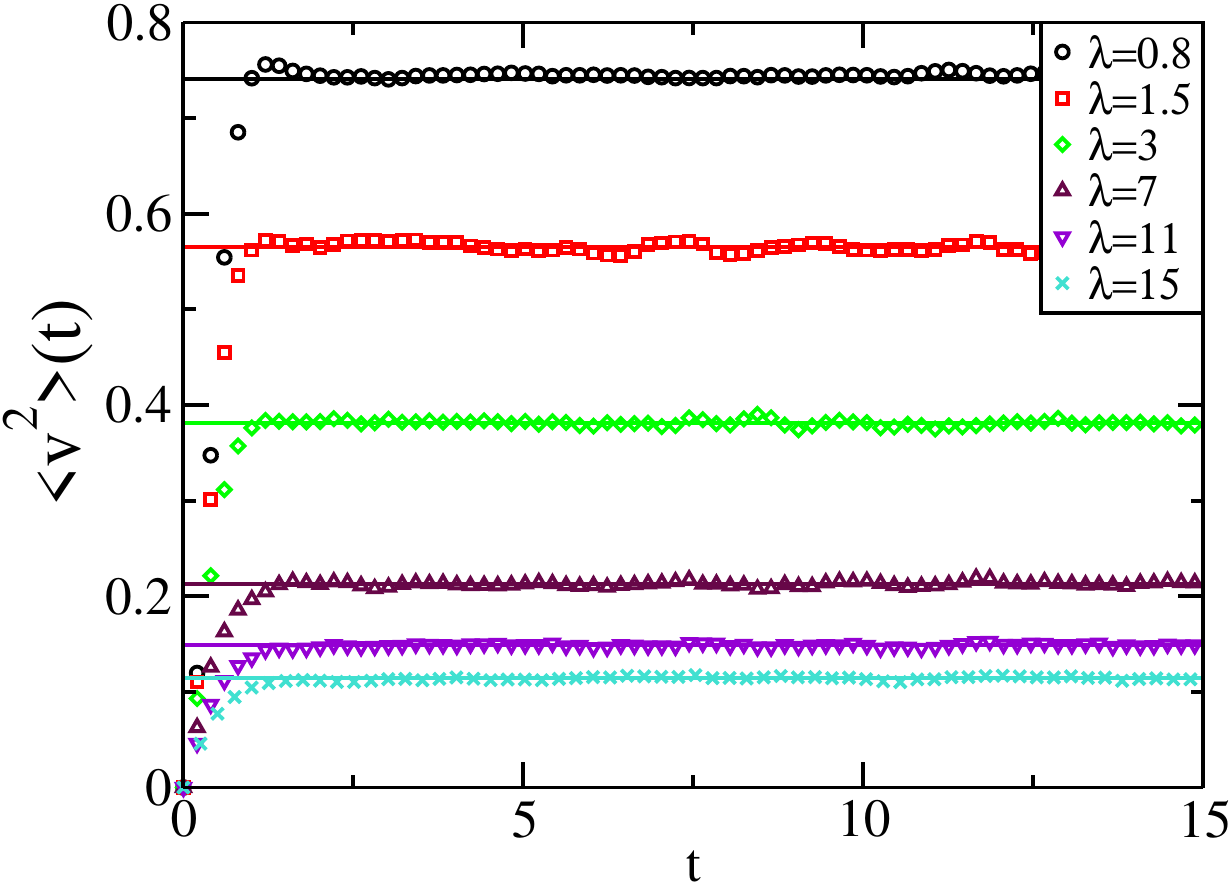}
\caption{\label{fig2} Plot of mean-squared velocity, $\langle v^2\rangle(t)$ vs. $t$ for $\Sigma=2$ and various values of $\lambda$, as mentioned. The data points represent numerical results, while the solid lines show the steady-state values of mean-squared velocity $\langle v^2\rangle_s$ computed from Eq.~(\ref{apeqn34}). Data points and solid lines of the same color correspond to the same values of $\Sigma$ and $\lambda$.}
\end{figure}

Figure~\ref{fig2} shows the plot of $\langle v^2\rangle(t)$ as a function of $t$ for $\Sigma=2$ and different values of $\lambda$, as indicated. After an early transient, $\langle v^2\rangle(t)$ reaches a steady-state value for all cases. We then numerically evaluate steady-state mean-squared velocity $\langle v^2\rangle_s$ by solving the following equation:
\begin{eqnarray}
\label{apeqn34}
\langle v^2\rangle_s = \frac{\int_{-v_m}^{+v_m} v^2 W_s(v) dv}{\int_{-v_m}^{+v_m}W_s(v) dv} .
\end{eqnarray}
Clearly, in the steady state, the numerical values of $\langle v^2\rangle(t)$ for long times agree with the numerically computed $\langle v^2\rangle_s$ for all cases. Smaller values of $\lambda$ lead to higher $\langle v^2\rangle_s$, as lower $\lambda$ allows the particle to reach the maximum possible velocities, $v_m$, for a fixed $\Sigma$. 

\begin{figure}
\centering
\includegraphics*[width=0.9\textwidth]{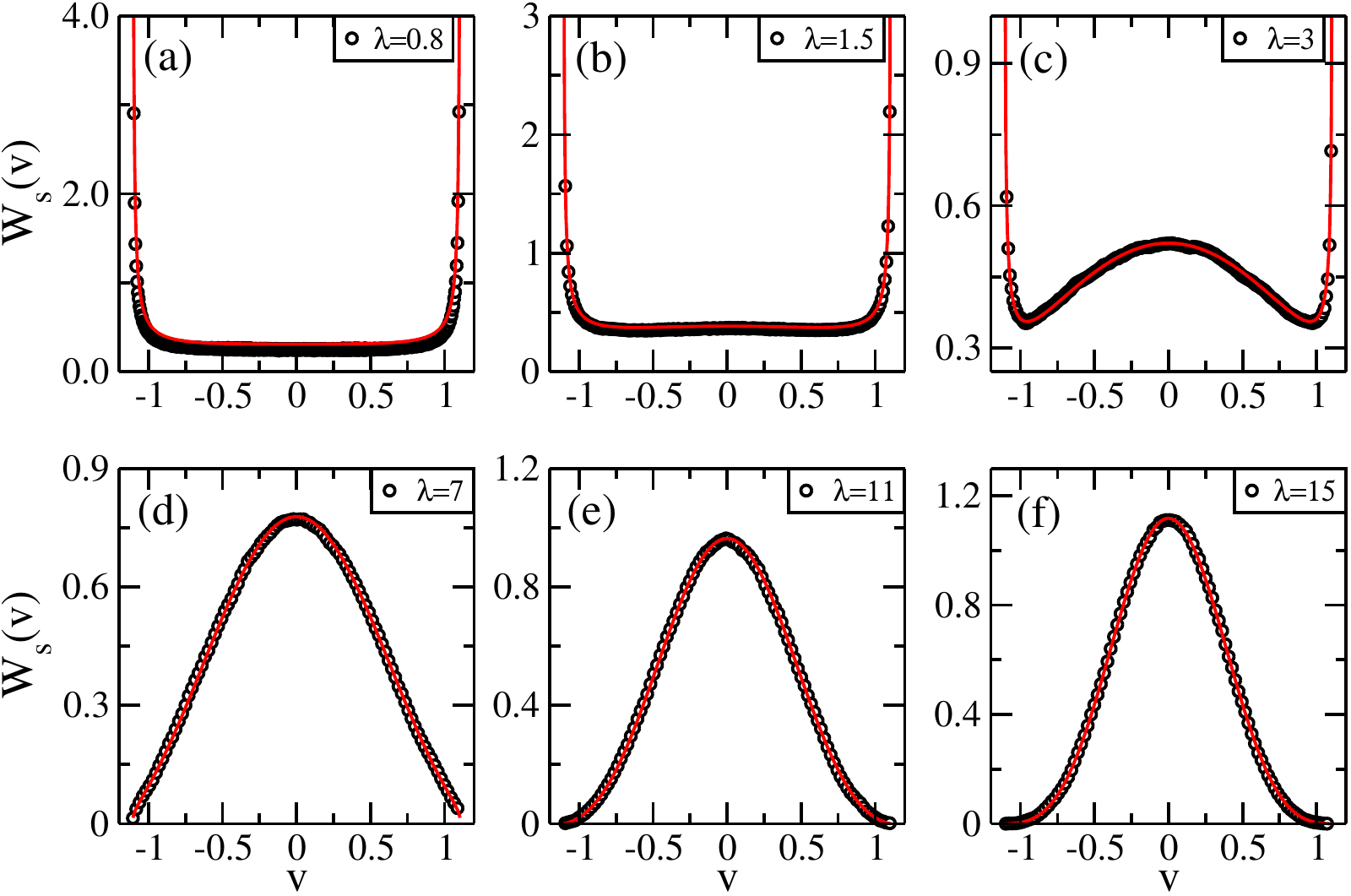}
\caption{\label{fig3} Plot of the normalized steady-state velocity distribution function $W_s(v)$ for $\Sigma=2$ and different values of $\lambda$: (a) $\lambda=0.8$, (b) $\lambda=1.5$, (c) $\lambda=3$, (d) $\lambda=7$, (e) $\lambda=11$, and (f) $\lambda=15$. The black circles correspond to numerical data, while the solid red lines represent the normalized form of $W_s(v)$ as given in Eq.~(\ref{apeqn18}).}
\end{figure}

Next, we compute the normalized steady-state velocity distribution $W_s(v)$ by collecting particle velocities and constructing a histogram from the collected data. Velocities were sampled over the time interval $t \in (100, 200)$ for all $10^5$ trajectories. The particle dynamics reach the steady state within this interval. Since $dt = 0.001$, each histogram was constructed using $10^{10}$ data points. In Fig.~\ref{fig3}, we plot the numerically computed and analytical (normalized) $W_s(v)$ for $\Sigma = 2$ and various values of $\lambda$, as indicated. The numerical and analytical results are nearly identical, confirming excellent agreement. This consistency holds for other values of $\Sigma$ and $\lambda$ as well.

\begin{figure}
\centering
\includegraphics*[width=0.9\textwidth]{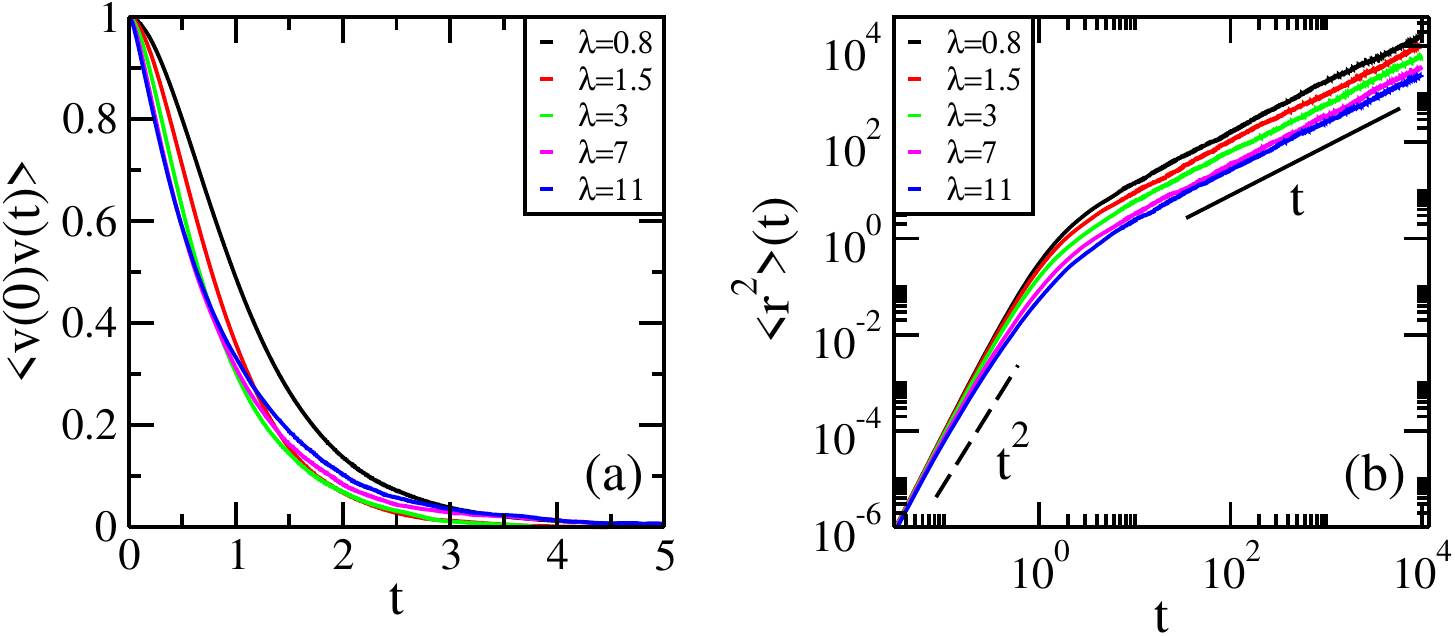}
\caption{\label{fig4} (a) Plot of the velocity autocorrelation function $\langle v(0)v(t)\rangle$ vs. $t$  for $\Sigma=2$ and various values of $\lambda$, as indicated. (b) Plot of the mean-squared displacement $\langle r^2\rangle(t)$ vs. $t$ on a log-log scale for the same values of $\Sigma$ and $\lambda$ as used in(a). The dashed line labeled $t^2$ and the solid line labeled $t$ represent the ballistic and diffusive regimes, respectively.}
\end{figure}

We numerically compute the velocity auto-correlation function $C(t)=\langle v(0)v(t)\rangle$ in the steady-state and the mean-squared displacement $\langle r^2\rangle(t)$ to investigate the transport properties of the system. Figure~\ref{fig4}(a) shows the plot of $C(t)$ vs. $t$ for $\Sigma=2$ and various values of $\lambda$, as indicated. It is evident that the numerical value of $C(t)$ decays to zero for all values of $\lambda$, indicating that the particle's velocity becomes statistically independent of its initial value. This implies that the particle will exhibit normal diffusive motion in the long-time limit. From Eq.~(\ref{apeqn26}), it follows that the particle has a finite effective diffusion coefficient, $D_{\rm eff}$. This conclusion is further supported by examining the variation of $\langle r^2\rangle(t)$ as a function of $t$. In Fig.~\ref{fig4}(b), we plot $\langle r^2\rangle(t)$ vs. $t$ on a log-log scale for $\Sigma=2$ and different values of $\lambda$, as mentioned. At early times, $\langle r^2\rangle(t)\sim t^2$, which corresponds to the ballistic regime. This superdiffusive behavior naturally arises from the initial deterministic motion of the particle before viscous effects dominate and the particle’s velocity becomes fully randomized by the active force. However, at later times, the particles exhibit diffusive motion, where $\langle r^2\rangle(t)\sim t$. The finite slope of the $\langle r^2\rangle(t)$ vs. $t$ curve (or the finite intercept of the $\langle r^2\rangle(t)$ vs. $t$ graph on a log-log scale) confirms that the particle has a finite $D_{\rm eff}$ for all allowed values of $\Sigma$ and $\lambda$.

\begin{figure}
\centering
\includegraphics*[width=0.96\textwidth]{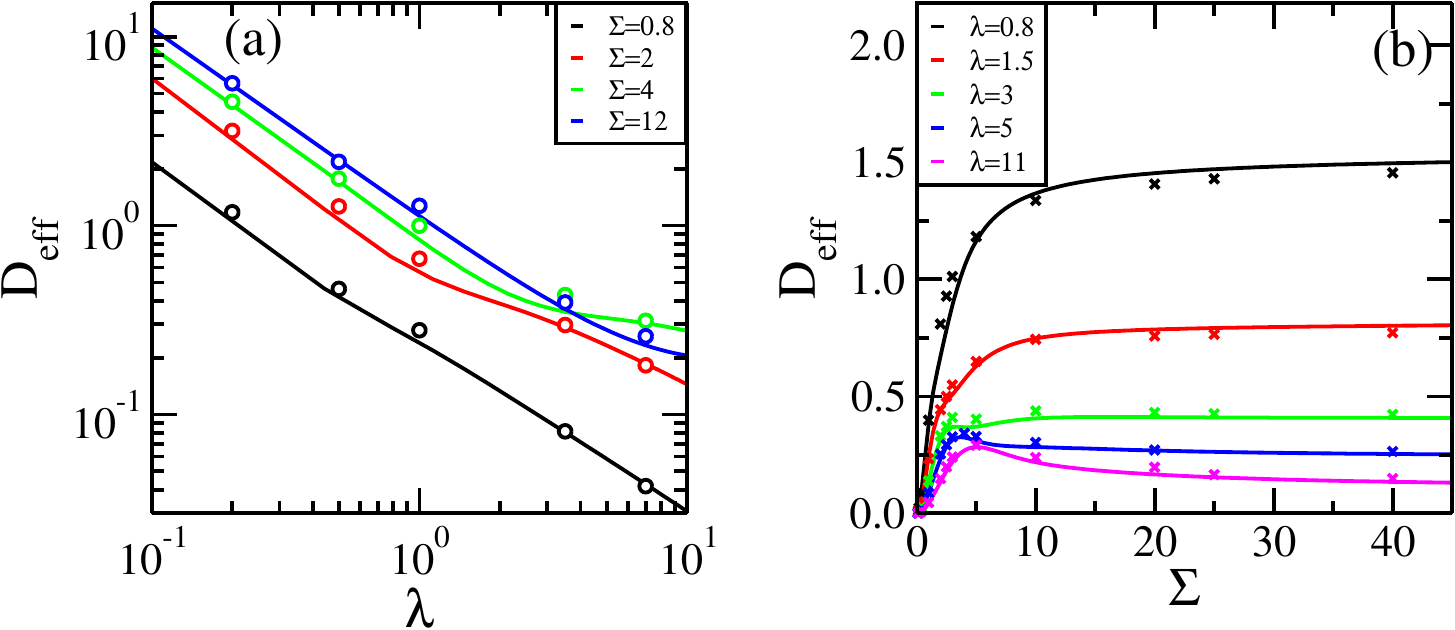}
\caption{\label{fig5} Plot of the effective diffusion coefficient $D_{\rm eff}$ for different $\Sigma$ and $\lambda$. (a) Log-log plot of $D_{\rm eff}$ vs. $\lambda$ for various values of $\Sigma$ and (b) plot of $D_{\rm eff}$ vs. $\Sigma$ for different values of $\lambda$, as mentioned. The numerical data obtained by integrating the velocity autocorrelation function are represented by points, while the lines correspond to $D_{\rm eff}$ computed numerically using Eq.~(\ref{apeqn27}). Lines and points of the same color represent $D_{\rm eff}$ for the same values of $\Sigma$ and $\lambda$.}
\end{figure}

Finally, we study the dependence of $D_{\rm eff}$ on the activity parameters $\Sigma$ and $\lambda$. First, we compute $D_{\rm eff}$ by integrating $C(t)$ over time (see Eq.~(\ref{apeqn26})) and also from the slope of the $\langle r^2\rangle(t)$ vs. $t$ curves shown in Fig.~\ref{fig4}(b). For a given $\Sigma$ and $\lambda$, the values of $D_{\rm eff}$ obtained using both methods are consistent with each other. Next, we calculate $D_{\rm eff}$ using its quadrature expression in Eq.~(\ref{apeqn27}) for various values of $\Sigma$ and $\lambda$. A detailed numerical method for evaluating the quadrature expression is provided in Ref.~\cite{B10}. Figure~\ref{fig5} shows plots of (a) $D_{\rm eff}$ versus $\lambda$ for a range of $\Sigma$ values on a log-log scale and (b) $D_{\rm eff}$ versus $\Sigma$ for a range of $\lambda$ values, using the methods mentioned earlier. In both instances, the $D_{\rm eff}$ computed from the quadrature expression closely matches that obtained from the slope of the $\langle r^2\rangle(t)$ vs. $t$ curve. In Fig.~\ref{fig5}(a), $D_{\rm eff}$ consistently decreases nonlinearly with increasing $\lambda$ for a fixed $\Sigma$. This occurs because increasing $\lambda$ drives the particle away from the maximum allowed velocity $v_m$ at a given $\Sigma$, causing it to fluctuate around $v = 0$. Consequently, $\langle r^2\rangle(t)$, and hence  $D_{\rm eff}$, decreases at a fixed $t$ and $\Sigma$ as $\lambda$ increases. In Fig.~5(b), for $\lambda = 0.8$, $1.5$, and $3$, $D_{\rm eff}$ gradually increases with increasing $\Sigma$ in the small-$\Sigma$ regime and then saturates in the large-$\Sigma$ regime. In contrast, for $\lambda = 5$ and $11$, $D_{\rm eff}$ exhibits a prominent peak at intermediate values of $\Sigma$, differing from the trend observed at the lower $\lambda$ values. We numerically identify the transition point $\lambda_c$ at which a prominent peak in $D_{\rm eff}$ emerges for the first time, yielding $\lambda_c \approx 3.5$ for $\Sigma = 2$. Additionally, we numerically observe that $W_s(0) \approx W_s(\pm v_m)$ at $\lambda_c$. For $\lambda > \lambda_c$, we find $W_s(0) > W_s(\pm v_m)$ [cf. Fig.~\ref{fig1}(b)], indicating that the particle velocity predominantly fluctuates around zero, leading to a decrease in $D_{\rm eff}$ with increasing $\lambda$ beyond $\lambda_c$. In contrast, for $\lambda < \lambda_c$, where $W_s(0) < W_s(\pm v_m)$[cf. Fig.~\ref{fig3}(c)], the velocity tends to remain near $\pm v_m$, resulting in an increase in $D_{\rm eff}$ as $\lambda$ decreases below $\lambda_c$. Finally, in the large-$\Sigma$ regime, $D_{\rm eff}$ becomes independent of $\Sigma$ for a fixed $\lambda$. This can be understood as follows: for large $\Sigma$, the velocity bound $v_m = \pm \tan^{-1} \Sigma$ approaches $\pm \pi/2$, making $D_{\rm eff}$ effectively independent of $\Sigma$ for $\lambda \lesssim \lambda_3 = 2(1 + \Sigma^2)$. For $\lambda > \lambda_3$, the particle velocity does not reach $v_m$, which again leads to $\Sigma$-independent behavior of $D_{\rm eff}$.

\section{Robustness of Transitions in $W_s(v)$}\label{sec4}
\begin{figure}
\centering
\includegraphics*[width=0.55\textwidth]{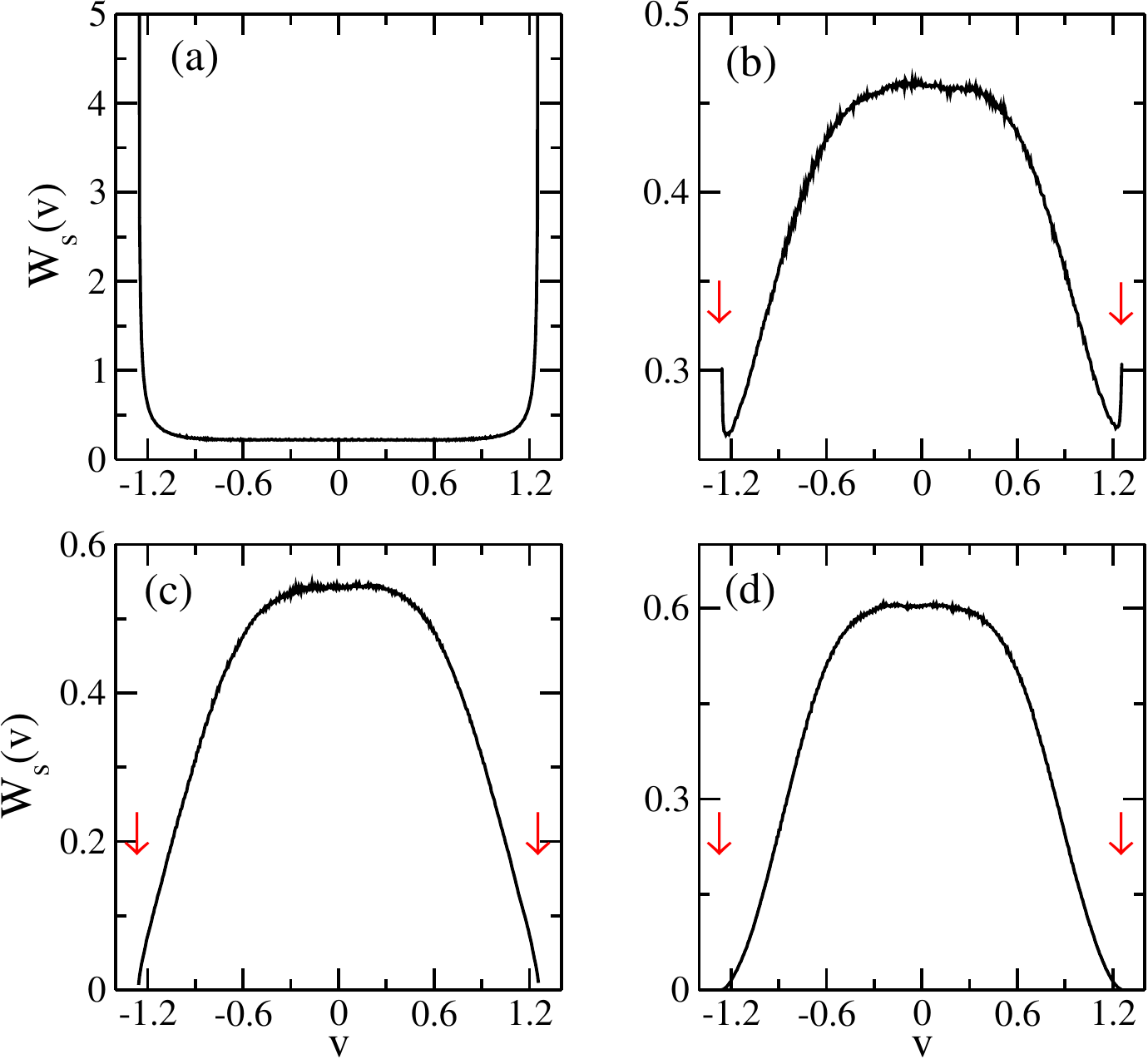}
\caption{\label{fig6} Plot of the normalized steady-state velocity distribution function $W_s(v)$ for $f(v) \sim v^3$ at $\Sigma = 2$ and various values of $\lambda$: (a) $\lambda = 0.8$, (b) $\lambda = 4.5$, (c) $\lambda = 8$, and (d) $\lambda = 12$, as indicated. The red arrows in panels (b)–(d) indicate the boundaries of the velocity domain.}
\end{figure}

In Section~\ref{2b}, we observed that $W_s(v)$ exhibits multiple transitions in the vicinity of $v = \pm v_m$, influenced by the active force parameters $\Sigma$ and $\lambda$, when the viscous force is modeled as $f(v) \sim \tan(v)$. This observation prompts a key question: Are these transitions universal across different choices of $f(v)$ representing the shear-thickening medium? To explore this, we consider $f(v) \sim v^3$, which also describes the viscous force in a shear-thickening medium, as $f(v)$ increases with increasing $v$. In this case, the scaled Langevin equation for the particle's motion becomes
\begin{eqnarray}
\label{apeqn41}
\frac{dv}{dt} = -v^3 + \eta(t).
\end{eqnarray}
Here, all variables have the usual meanings as described in Sec.~\ref{sec2}. In this case, the maximum velocity the particle can attain is $v_m = \pm\Sigma^{1/3}$. We numerically solve Eq.~(\ref{apeqn41}) using the Euler discretization method with a time step of $dt = 0.001$ to compute the time-dependent velocity of the particle, $v(t)$. The system is equilibrated up to $t = 100$, after which $v(t)$ is recorded over the interval $t \in (100, 200)$ for $10^5$ independent trajectories to compute $W_s(v)$. Figure~\ref{fig6} shows the plots of $W_s(v)$ for $f(v) \sim v^3$ at $\Sigma=2$ and four different values of $\lambda$, as mentioned. Clearly, $W_s(v)$ for $f(v) \sim v^3$ exhibits all the possible behaviors near $\pm v_m$ that were observed for $f(v) \sim \tan(v)$. This confirms that the behavior of $W_s(v)$ near the maximum attainable velocity is independent of the specific form of $f(v)$, highlighting the robustness of the transitions near $\pm v_m$. However, the $\Sigma$–$\lambda$ relationships that determine the transition points may depend on the form of $f(v)$.

\section{Summary and Discussion}\label{sec5}
In summary, we have studied the dynamics of an athermal inertial run-and-tumble particle navigating a shear-thickening medium in one dimension ($d=1$). The nonlinear viscous force exerted on the particle by the medium is characterized by the relation $f(v) \sim \tan(v)$. A symmetric dichotomous noise, characterized by an amplitude $\Sigma$ and a flipping rate $\lambda$, emulates the behavior of a run-and-tumble particle. Starting from the non-dimensional form of the Langevin equation governing the particle's motion, we derive the Fokker–Planck (FP) equation for the time-dependent probability distribution $W_{\pm\Sigma}(v, t)$. Next, we analytically obtain the steady-state velocity distribution function $W_s(v)$. Our analysis reveals that $W_s(v)$’s shape depends on both $\Sigma$ and $\lambda$. For a fixed $\Sigma$, $W_s(v)$ exhibits distinct transitions as $\lambda$ varies: (a) For $\lambda < 1$, $W_s(v)$ has a minimum at $v = 0$ and diverges at $v = \pm \tan^{-1}\Sigma$; (b) For $1 < \lambda < 1 + \Sigma^2$, a local maximum emerges at $v = 0$, minima appear at $v = \pm \tan^{-1}\sqrt{\lambda - 1}$, and $W_s(v)$ continues to diverge at $v = \pm \tan^{-1}\Sigma$; (c) For $1 + \Sigma^2 < \lambda < 2(1 + \Sigma^2)$, $v = 0$ becomes the sole maximum, and $W_s(v)$ approaches zero at $v = \pm \tan^{-1}\Sigma$ with a cusp-like profile; (d) For $\lambda > 2(1 + \Sigma^2)$, both $W_s(v)$ and its derivative $W_s^\prime(v)$ approach zero at $v = \pm \tan^{-1}\Sigma$, while $v = 0$ remains the maximum. These findings highlight the intricate interplay between activity parameters and the resulting velocity distribution in such systems. Here, we want to emphasize that the transitions observed near $v = \pm v_{m}$ are independent of the specific form of the functions used to model the shear-thickening behavior of the medium. However, the $\Sigma$-$\lambda$ relationship that determines the transition points may depend on the specific form of the function modeling the shear-thickening behavior of the medium. Additionally, we obtain an exact quadrature expression for $D_{\rm eff}$. However, determining the exact analytical form of dependence of $D_{\rm eff}$ on $\Sigma$ and $\lambda$ remains challenging due to non-existence of the closed form of the integrals.

We numerically integrate Langevin’s equation of motion for a particle to determine its time-dependent velocity, $v(t)$, and position, $r(t)$. Subsequently, we compute the mean-squared velocity, $\langle v^2\rangle(t)$, as a function of time. At sufficiently long times, $\langle v^2\rangle(t)$ converges to a steady-state value, $\langle v^2\rangle_s$, which is consistent with the steady-state value derived numerically from the distribution $W_s(v)$. Furthermore, for fixed values of $\Sigma$ and $\lambda$, the numerically obtained normalized steady-state distribution $W_s(v)$ shows good agreement with its analytical form.

To evaluate the effective diffusion coefficient, $D_{\rm eff}$, we calculate the velocity autocorrelation function, $C(t)=\langle v(0)v(t)\rangle$, and the mean-squared displacement, $\langle r^2\rangle(t)$, for various combinations of $\Sigma$ and $\lambda$. At large times $t$, the relationship $\langle r^2\rangle(t)\sim t$ emerges, indicative of diffusive particle behavior in the long-time limit. For fixed $\Sigma$ and $\lambda$, the $D_{\rm eff}$ obtained by integrating $C(t)$ over time aligns with both the value extracted from the slope of the $\langle r^2\rangle(t)$ vs. $t$ plot in the diffusive regime and the analytical prediction. For a fixed $\Sigma$, $D_{\rm eff}$ decreases nonlinearly as $\lambda$ increases. Additionally, for a constant $\lambda$, $D_{\rm eff}$ exhibits independence from $\Sigma$ in the large-$\Sigma$ regime.

We believe that our study offers a fundamental understanding of the steady-state velocity distribution and transport properties of athermal inertial run-and-tumble particles moving through a shear-thickening medium in $d=1$. Extending this analysis to higher dimensions and investigating the influence of spatial dimensionality on various dynamical quantities presents an exciting research direction. A promising avenue for future work would be to explore the time-dependent and steady-state transport properties of different active systems in shear-thickening media, such as inertial active Ornstein-Uhlenbeck particles, run-and-tumble particles in harmonic traps, and active particles driven by external fields. Moreover, recent advances in robotics have enabled the development of self-propelled micro- and nanorobots designed to navigate through non-Newtonian media, including biological fluids like mucus and blood, for efficient drug delivery. Since these robots have finite mass, our results for $W_s(v)$ and $D_{\rm eff}$ could be directly applied to describe their motion. This is just one example from applied sciences where our study could have practical relevance. We hope that our findings will motivate further experimental research across a broad range of scenarios, connecting biological physics with engineering applications.

\ \\
\noindent{\bf Acknowledgments:} SH and SM acknowledge financial support from IISER Mohali through a Senior Research Fellowship. PD acknowledges financial support from SERB, India through a start-up research grant (SRG/2022/000105).

\ \\
\noindent{\bf Conflict of interest:} The authors have no conflicts to disclose.

\end{document}